\documentclass[twocolumn,english,prl,amssymb,aps,superscriptaddress,showpacs,twocolumn,amsmath,showkeys,floatfix]{revtex4-1}
\usepackage[T1]{fontenc}
\usepackage[latin9]{inputenc}
\usepackage{geometry}
\usepackage[active]{srcltx}
\usepackage{textcomp}
\usepackage{amsmath}
\usepackage{graphicx}
\usepackage{esint}

\makeatletter
\usepackage{babel}

\usepackage{babel}

\makeatother

\usepackage{babel}
\begin{document}

\title{Excess noise factor in non-Hermitian mode-locked lasers}
\author{Oriane Leli\`evre}
\affiliation{Laboratoire Aim\'e Cotton, Universit\'e Paris-Sud, ENS Paris-Saclay, CNRS, Universit\'e Paris-Saclay, 91405 Orsay, France}
\affiliation {Thales Research and Technology, 91120 Palaiseau, France}
\author{Khadija Abensar}\author{Ghaya Baili }
\author{Vincent Crozatier }
\affiliation {Thales Research and Technology, 91120 Palaiseau, France}
\author{Fabienne Goldfarb }
\affiliation{Laboratoire Aim\'e Cotton, Universit\'e Paris-Sud, ENS Paris-Saclay,
CNRS, Universit\'e Paris-Saclay, 91405 Orsay, France}
\author{Fabien Bretenaker}
\affiliation{Laboratoire Aim\'e Cotton, Universit\'e Paris-Sud, ENS Paris-Saclay,
CNRS, Universit\'e Paris-Saclay, 91405 Orsay, France}
\affiliation{Light and Matter Physics Group, Raman Research Institute, Bangalore 560080, India}

\begin{abstract} We predict and observe the appearance of an excess noise factor due to the non-Hermiticity of the modes of two actively harmonically mode-locked lasers. The non uniform distribution of gain and losses in the ring cavities of these lasers causes a non-orthogonality of the longitudinal modes, which is responsible for an increase of the rate of spontaneous emission falling in the lasing pulsed mode. This leads to an increase of the phase noise of the pulse train generated by these lasers by more than 20 dB at some offset frequencies. This degradation can be prohibitive for the use of these lasers for several applications such as the generation of high spectral purity RF and low jitter clock signals.
\end{abstract}

\pacs{42.55.Ah, 42.50.Lc, 42.62.Eh, 42.60.Fc}

\maketitle
The laser linewidth is usually considered to be given by the famous Schawlow-Townes formula \cite{Schawlow1958}, which can be classically viewed as due to phase diffusion of the laser field induced by spontaneous emission \cite{Jacobs1979}. This formula, originally obtained in an ideal case, has since been completed by many correction factors, due for example to incomplete population inversion \cite{Siegman1968, Manes1971}, internal cavity losses \cite{SiegmanLivre}, so-called bad cavity regime \cite{VanExter1995}, or phase-intensity coupling, which is particularly large for semiconductor active media \cite{Henry1982,Henry1986}. Another discrepancy with respect to the original Schawlow-Townes formula happens when the eigenmodes of the laser cavity are not orthogonal \cite{Siegman1979}. In such  non-Hermitian resonators \cite{Siegman1989a,Siegman1989b}, each mode sees a fraction of the spontaneous emission that falls into the other modes on top of its own share of spontaneous emission \cite{Mussche1990,New1995,Grangier1998}. This causes an increase of the laser linewidth, the so-called \textit{excess noise factor} (ENF). This effect has manifested itself through the non-orthogonality of the transverse modes of geometrically unstable \cite{Cheng1996,vanEijkelenborg1996} and stable resonators \cite{Brunel1997}, the non-orthogonality of the guided modes of gain-guided semiconductor lasers \cite{Petermann1979}, the non-orthogonality of the longitudinal modes of lasers exhibiting large losses \cite{Hamel1989,Hamel1990}, and the non-orthogonality of the two polarization states in the laser \cite{VanderLee1997,Emile1998}. In all these cases, the non-Hermiticity of the cavity leads to the cavity eigenmodes being no longer orthogonal but are rather biorthogonal to a set of adjoint modes \cite{SiegmanLivre, Siegman1979}. Then, when the spontaneous emission is projected on one laser mode in order to derive the Langevin force driving the fluctuations of this mode, one obtains a coefficient larger than 1, which is precisely the ENF

Most experimental investigations of the laser ENF were performed with continuous wave lasers exhibiting only one or two oscillating modes. On the contrary, mode-locked lasers can also be seen as optical frequency combs, exhibiting a large number of oscillating modes. Since they play a very important role in recent advances in physics and metrology such as high resolution spectroscopy \cite{Hansch2006}, multimode quantum optics \cite{Gerke2016}, frequency metrology \cite{Bernhardt2010}, and development of ultra-low phase noise RF oscillators \cite{Fortier2011,Bouchand2016,Xie2017} and optically carried RF clocks \cite{Mandridis2010}, the control of their RF and optical phase noise is of crucial importance. In such a strongly multimode mode-locked laser, the Schawlow-Townes linewidth for each optical frequency is the same as the one for a single-frequency laser, with an oscillating mode power equal to the sum of the powers of all the modes of the mode-locked laser \cite{Ho1985}. This somewhat counterintuitive feature comes from the fact that a mode-locked laser can also be seen as a single-mode laser when one quantizes the field along pulsed modes, i. e. photon wave packets, rather than traveling modes \cite{GAF}.  
However, most mode-locked lasers contain several intracavity elements: for example, the cavity of an actively mode-locked laser must contain a modulator to which an RF signal is applied at an harmonic of the laser free spectral range \cite{SiegmanLivre}. Such intracavity elements exhibit strong losses and one can thus wonder whether the longitudinal modes are still orthogonal in such a cavity, and whether the possible non-Hermiticity of the cavity may lead to an excess optical phase noise, that would be converted into an extra RF phase noise on the beatnote between the modes.

\begin{figure}[htbp]
\centering
\includegraphics[width=0.99\columnwidth]{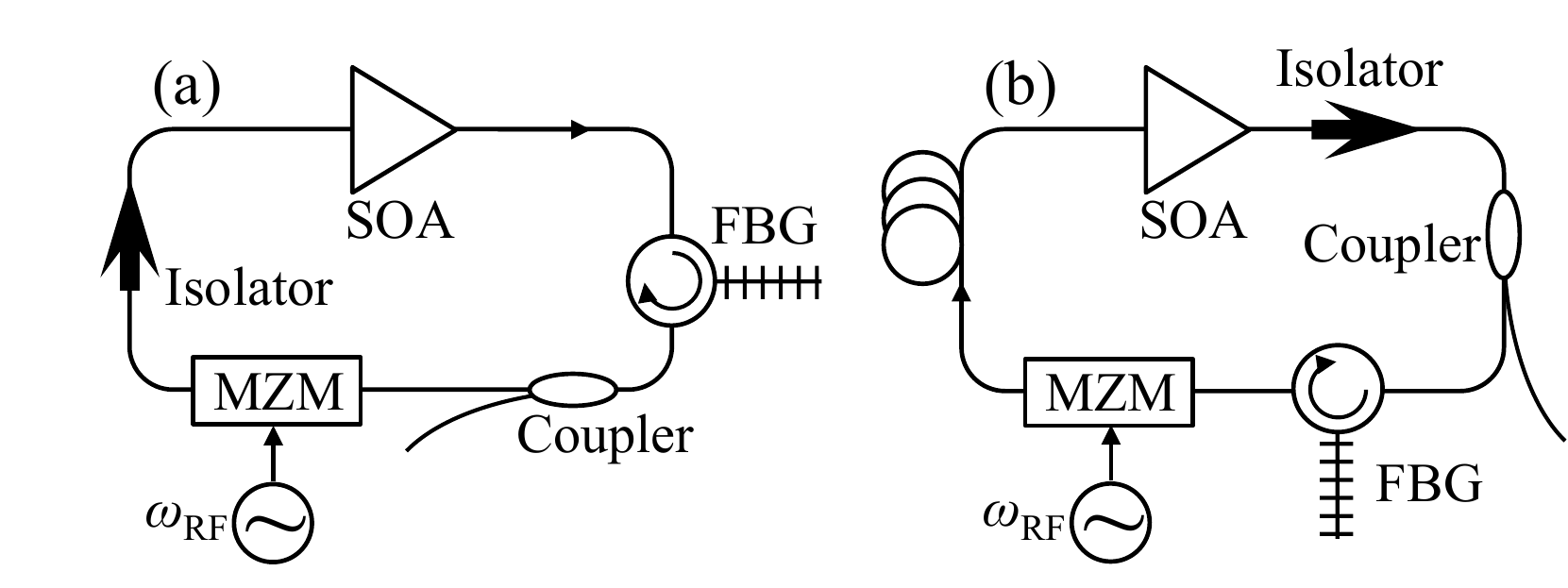}
\caption{AHMLL oscillating at (a) 800 nm and (b) 1550 nm. SOA: semiconductor optical amplifier; MZM: Mach-Zehnder modulator; FBG: fiber Bragg grating.}
\label{Fig01}
\end{figure}

The kind of laser that we consider here is schematized in Fig.\;\ref{Fig01}. It is an Actively Harmonically Mode-Locked Laser (AHMLL) based on a fibered cavity. Such lasers are being actively considered for the development of low-jitter optical pulsed sources \cite{Mandridis2010} and coupled optoelectronic oscillators, which are among the quietest RF sources in the 10-100~GHz domain \cite{Yao1997,Matsko2009, Matsko2013,Ly2018}. The active media are Semiconductor Optical Amplifiers (SOAs) providing gain either at 800 or at 1550\;nm, corresponding to the cavities of Figs. \ref{Fig01}(a,b), respectively. In both cases, harmonic mode-locking is obtained through RF modulation of the losses by an integrated Mach-Zehnder intensity modulator (MZM) driven at  frequency $\omega_{\mathrm{RF}}$ close to the $p^{\mathrm{th}}$ harmonic of the cavity free spectral range $\Delta$. The cavity also contains an optical filter to control the optical spectrum, an isolator to force oscillation in a single direction and a fiber coupler to extract some optical power for detection and analysis.

When the train of pulses emitted by such lasers is detected by a fast photodiode, one obtains an RF signal at $\omega_{\mathrm{RF}}$ whose phase noise contains two contributions due to i) the noise of the RF source feeding the modulator and ii) spontaneous emission noise. This leads to the following power spectral density of the phase noise of the RF signal at $\omega_{\mathrm{RF}}$ \cite{Hjelme1992, Quinlan2008}:
\begin{eqnarray}
S_{\varphi}(\omega)=\sum_{m=-\infty}^{+\infty}\frac{1}{(\omega-m\Delta)^2+\Gamma^2}\left[\Gamma^2S_{\mathrm{RF}}(\omega)\right. \nonumber\\
\left.+\frac{2\sqrt{2}ø\Delta\omega_{\mathrm{ST}}}{N}\;\times \left(1+\frac{2N^2\delta^2}{(\omega-m\Delta)^2+4\Gamma^2}\right)\right],\label{eq01}
\end{eqnarray}
where $\Gamma$ is the mode-locked laser characteristic filter angular frequency, $N$ is proportional to the number of lasing modes in the sense that $N \omega_{\mathrm{RF}}$ is the $1/e^2$ half-width of the laser spectrum,  $S_{\mathrm{RF}}(\omega)$ is the power spectral density of the phase noise of the RF source driving the modulator, $\delta=\omega_{\mathrm{RF}}-p\Delta$ is the detuning between the RF source and the relevant harmonic of the free spectral range. Since the laser is harmonically mode-locked, the repetition frequency $\omega_{\mathrm{RF}}$ is an integer multiple of $\Delta$ and several frequency combs coexist in the laser. Equation (\ref{eq01}) supposes that the different combs are independent. Finally, $\Delta\omega_{\mathrm{ST}}$ is the  Schawlow-Townes linewidth given by:
\begin{equation}
\Delta\omega_{\mathrm{ST}}=(1+\alpha^2)\frac{h\nu}{2\tau_{\mathrm{cav}}^2 P_{\mathrm{out}}}\frac{T_{\mathrm{out}}}{1-\Upsilon}\ .\label{eq02}
\end{equation}
$h\nu$ is the energy of a photon, $\alpha$ the Henry factor \cite{Henry1986} of the SOA, $\tau_{\mathrm{cav}}=2\pi/\Delta\Upsilon$ the cavity photon lifetime, $\Upsilon$ the total losses per round-trip, $P_{\mathrm{out}}$ the laser output power, and $T_{\mathrm{out}}$ the output coupler transmission.

We used eqs. (\ref{eq01}) and (\ref{eq02}) to analyze the RF phase noise spectra of two different AHMLLs, operating respectively at 800\;nm and 1550\;nm. These two lasers also differ their repetition rates (4 GHz vs. 10 GHz), their intra-cavity fiber lengths (15 m vs. 100 m), and their intracavity filter bandwidths (1 nm vs. 2 nm), etc. 
\begin{figure}[htbp]
\centering
\includegraphics[width=0.8\columnwidth]{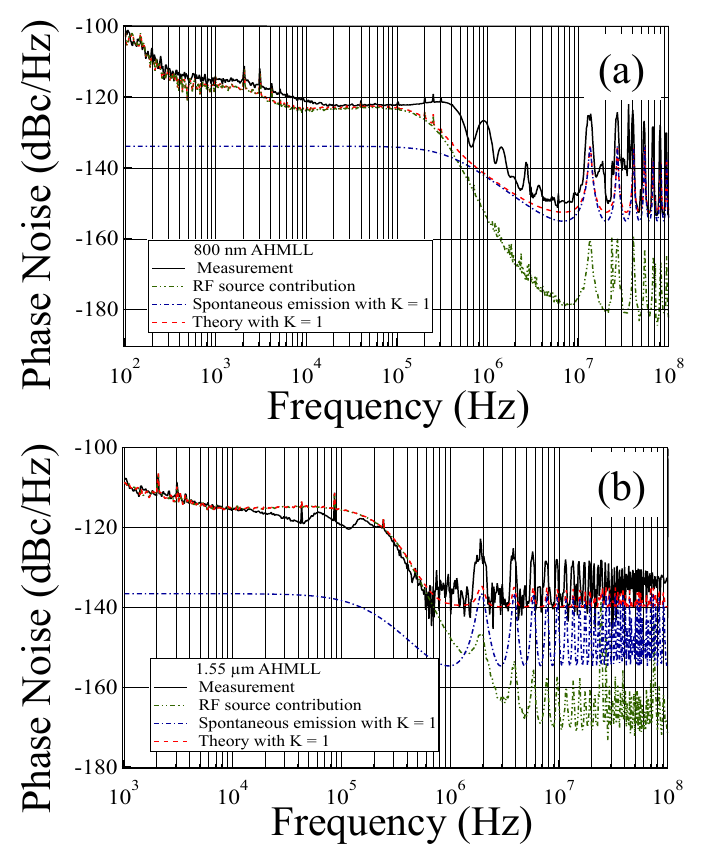}
\caption{(Color online) RF phase noise spectra of the AHMLL at (a) 800 nm and (b) 1550 nm. Thick full black lines: measurements; Dot-dot-dashed green lines: contribution of the RF source noise; Dash-dotted blue lines: contribution of spontaneous emission; Dashed red lines: theory adding both contributions plus the shot noise level equal to (a) $- 158\;\mathrm{dBc/Hz}$ and (b) $-141\;\mathrm{dBc/Hz}$. }
\label{Fig02}
\end{figure}

The results are reproduced in Fig.\;\ref{Fig02}. For both lasers, we compute the two components of the noise spectra given by eq.\,(\ref{eq01}) and compare them with experimental data. The calculation of the contribution of the RF source noise is based on a measurement of $S_{\mathrm{RF}}(\omega)$ at both frequencies (4 GHz and 10 GHz). The values of $N$ are obtained by fitting the measured optical spectra of the two lasers, leading to $N=3$ (resp. 4) for the 800\;nm (resp. 1550\;nm) laser. The values of $\Gamma$ ($2\pi\times15.0\times10^6\;\mathrm{rad/s}$ and $2\pi\times12.5\times10^6\;\mathrm{rad/s}$ respectively) are obtained by adjusting the experimental spectra. Finally the total losses $\Upsilon$ (equal to 16.6 and 25~dB respectively) are evaluated from the losses of the intra-cavity components and the output powers $P_{\mathrm{out}}$ (equal to 3.5 and 5~mW respectively) are measured. For both lasers, we have $T_{\mathrm{out}}=0.5$. With these values of the parameters and with $\alpha$ equal to 3 for the two lasers, we obtain $\Delta\omega_{\mathrm{ST}}=0.5\;\mathrm{rad/s}$ (resp. $\Delta\omega_{\mathrm{ST}}=0.1\;\mathrm{rad/s}$) for the 800\,nm (resp. 1550\,nm) laser. Finally, the shot noise level corresponding to the detected power is also added to the total phase noise of eq. (\ref{eq01}).

Comparison with experimental data (see Fig.\,\ref{Fig02}) shows that for both lasers, the noise induced by the RF source dominates at low frequencies (typically below 1 MHz) while spontaneous emission noise becomes dominant at higher frequencies. Moreover, these spectra clearly show that the noise is filtered by the cavity comb-like spectrum. The agreement of experiment with theory is very good for the low frequency part of the spectra. It is excellent in the case of the 1550~nm laser (see Fig. \ref{Fig02}(b)). It is slightly less good in the vicinity of the laser cut-off frequency in the case of the 800~nm laser (see Fig. \ref{Fig02}(a)), probably because of the occurrence of spectral interferences between different supermodes \cite{Quinlan2009}. On the contrary, comparison of experiment with theory shows that the contribution of spontaneous emission calculated from eq.\,(\ref{eq02}) is strongly underestimated for both lasers. Such an underestimation is reminiscent of what happens in single-frequency lasers containing large localized losses \cite{Hamel1989,Hamel1990} leading to a strong ENF due to the non-Hermiticity of the cavity.

\begin{figure}[htbp]
\centering
\includegraphics[width=0.99\columnwidth]{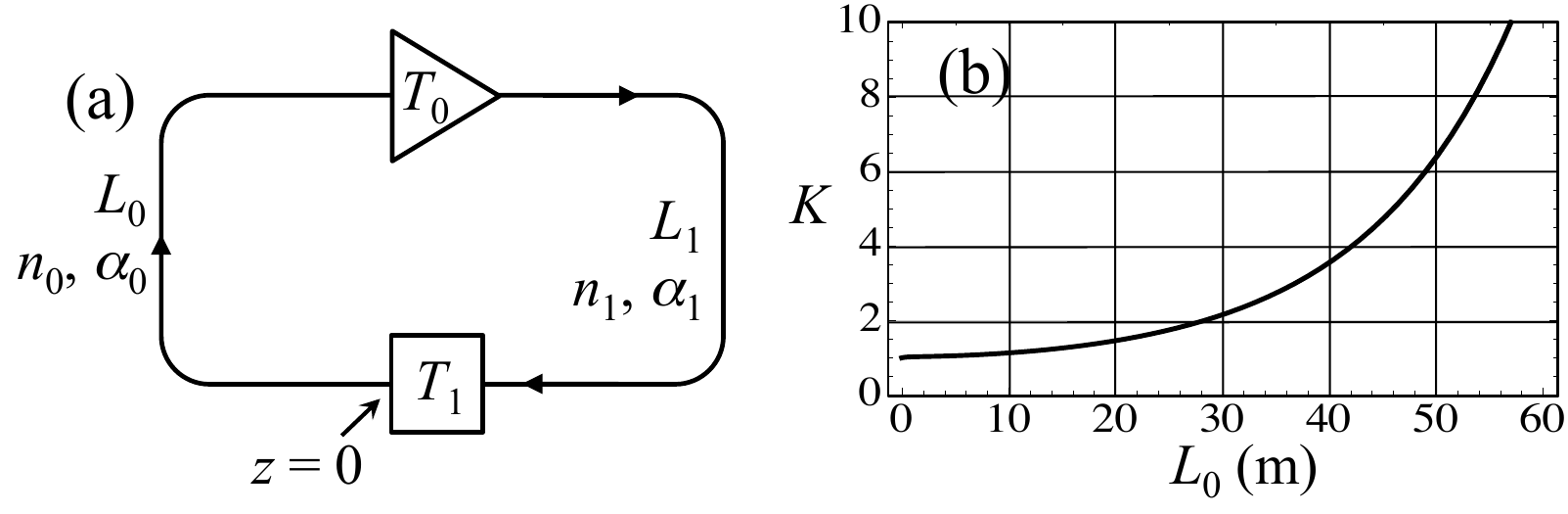}
\caption{(a) Cavity toy model containing losses and gain separated by two fiber sections (b) Corresponding evolution of the ENF $K$ versus $L_0$ with $L_1=1\;\mathrm{m}$, $\alpha_0=\alpha_1=0.1\;\mathrm{m}^{-1}$, and $T_1=0.8$. }
\label{Fig02N1}
\end{figure}

The origin of the ENF can be understood using the idealized cavity of Fig.\,\ref{Fig02N1}(a) as a toy model. It contains one gain element (transmission $T_0>1$) and one lossy element (transmission $T_1<1$), separated by two propagation segments, such as optical fibers, having lengths $L_i$, refractive indices $n_i$, and absorption coefficients $\alpha_i$ ($i=0,1$). The lasing mode ENF can be derived from the field distribution of this mode and of its adjoint mode. This adjoint mode is, in the most general case, obtained by solving the transposed cavity eigenequation \cite{SiegmanLivre,Siegman1989a}. In our present case, it is simply the mode propagating in the opposite direction \cite{Siegman1989a,Siegman1989b}, with gain sections replaced by losses, and vice-versa. Once the mode and its adjoint are obtained, we must normalize the cavity mode to 1, and the adjoint mode in such a way that its dot product with the cavity mode is also 1. The amount of spontaneous emission that falls in the laser mode is obtained by projecting the spontaneous emission onto the adjoint mode. The ENF is then shown to be given by the squared norm of this adjoint mode \cite{Siegman1989a,Siegman1989b,VanderLee1997,Alouini2002}. 

Starting from the origin $z=0$, where $z$ denotes the propagation distance inside the cavity, we write the mode amplitude $u(z)$ in each section of the cavity. For example, in the first cavity section $0\leq z\leq L_0$, it reads
\begin{equation}
u(z) = C \mathrm{e}^{(\mathrm{i}kn_0-\alpha_0/2)z}\ ,\label{eq03}
\end{equation}
where $k$ is the vacuum wavenumber and $C$ a normalization constant that will be derived later. Then the mode amplitude in the next section ($L_0\leq z\leq L_0+L_1$) is
\begin{equation}
u(z) = C\sqrt{T_0} \mathrm{e}^{(\mathrm{i}kn_0-\alpha_0/2)L_0}\times\mathrm{e}^{(\mathrm{i}kn_1-\alpha_1/2)(z-L_0)}\ .\label{eq04}
\end{equation}
The factor $C$ is determined by normalizing the mode to 
\begin{equation}
\int_0^{L_{\mathrm{tot}}}|u(z)|^2dz=1\ , \label{eq05}
\end{equation}
 with $L_{\mathrm{tot}}=L_0+L_1$  the total cavity length. Using eqs.\,(\ref{eq03}) and (\ref{eq04}), this leads to
 \begin{equation}
 |C|^2=\left(L_0^{\prime}+T_0\mathrm{e}^{-\alpha_0L_0}L_1^{\prime}\right)^{-1}\ ,\label{eq06}
 \end{equation}
with
\begin{equation}
L_m^{\prime}=\left(1-\mathrm{e}^{-\alpha_mL_m}\right)/\alpha_m\ ,\label{eq07}
\end{equation}
for $m=0,1$.

Since the cavity is not Hermitian, we need to calculate the adjoint mode $\varphi(z)$ of $u(z)$ by considering the reverse propagation direction. For example, in the last section ($L_0\leq z\leq L_{\mathrm{tot}}$), the adjoint mode amplitude reads
\begin{equation}
\varphi(z) = C' \sqrt{T_1}\mathrm{e}^{(\mathrm{i}kn_1-\alpha_1/2)(L_{\mathrm{tot}}-z)}\ ,\label{eq08}
\end{equation}
where $C'$ is a normalization constant. In the first cavity section ($0\leq z\leq L_0$), it becomes
\begin{equation}
\varphi(z) = C'\sqrt{T_0T_1} \mathrm{e}^{(\mathrm{i}kn_1-\alpha_1/2)L_1}\times\mathrm{e}^{(\mathrm{i}kn_0-\alpha_0/2)(L_0-z)}\ .\label{eq09}
\end{equation}
By definition of the adjoint mode, $C'$ must obey the following normalization:
\begin{equation}
\int_0^{L_{\mathrm{tot}}}u(z)\varphi(z)dz=1\ ,\label{eq10}
\end{equation}
which, using eqs.\,(\ref{eq03}), (\ref{eq04}), (\ref{eq08}), and (\ref{eq09}), leads to
\begin{equation}
CC'L_{\mathrm{tot}}=1\ .\label{eq11}
\end{equation}
Finally, the ENF for mode $u(z)$ is given by \cite{Siegman1989b}:
\begin{equation}
K=\int_0^{L_{\mathrm{tot}}}|\varphi(z)|^2dz>1\ ,\label{eq12}
\end{equation}
leading to:
\begin{equation}
K=\frac{T_1}{L_{\mathrm{tot}}^2}\left(L_1^{\prime}+T_0\mathrm{e}^{-\alpha_1 L_1}L_0^{\prime}\right)\left(L_0^{\prime}+T_0\mathrm{e}^{-\alpha_0 L_0}L_1^{\prime}\right)\ .\label{eq13}
\end{equation}

Figure \ref{Fig02N1}(b) shows a calculation of $K$ according to eq.\,(\ref{eq13}) for the toy cavity of Fig.\,\ref{Fig02N1}(a). We choose $L_1=1\;\mathrm{m}$, $\alpha_0=\alpha_1=0.1\;\mathrm{m}^{-1}$, and $T_1=0.8$. The value of $K$ is plotted as a function of $L_0$. For each value of $L_0$, $T_0$ is calculated in such a way that the round-trip gain exactly compensates for the losses, i. e.,  $T_0T_1=\exp(\alpha_0L_0+\alpha_1L_1)$. Figure\,\ref{Fig02N1}(b) shows that the ENF is close to 1 when the cavity is symmetric, i. e., $L_0\simeq L_1$ and $\alpha_0\simeq \alpha_1$, while it strongly exceeds one when the cavity asymmetry becomes strong. Since the cavities that we used to obtain the spectra of Fig.\,\ref{Fig02} exhibit strongly asymmetric distributions of gain and losses, it is thus quite likely that the discrepancy between experiments and theory should be attributed to the existence of a strong ENF $K$.

\begin{figure}[htbp]
\centering
\includegraphics[width=0.99\columnwidth]{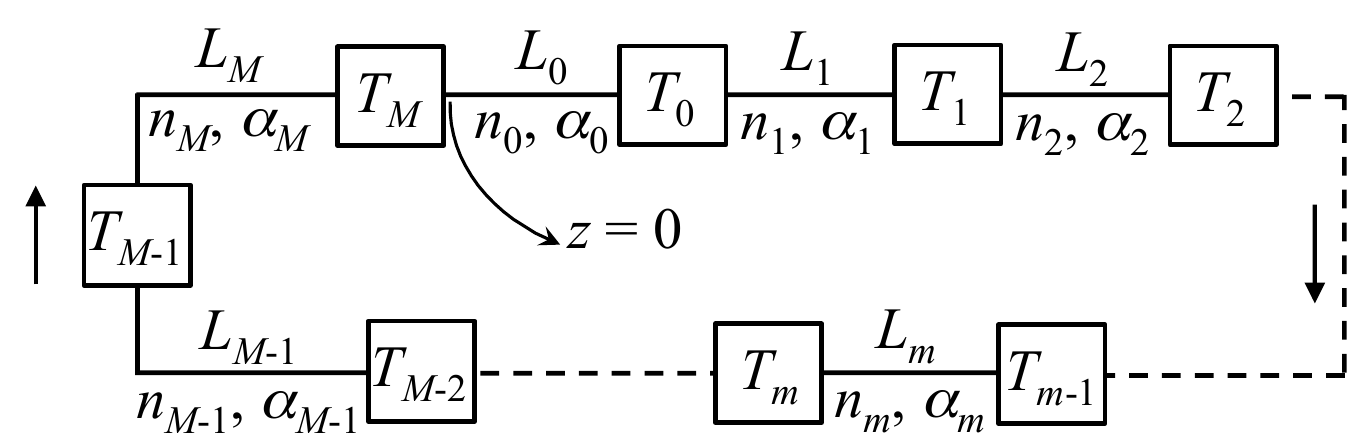}
\caption{Schematic cavity used to calculate the ENF $K$. }
\label{Fig03}
\end{figure}
To generalize the calculation of the longitudinal ENF to real cavities, we extend the preceding simple theory to a ring cavity containing $M+1$ localized lossy elements of transmissions $T_m$ ($m=0,\cdots,M$) separated by $M+1$ propagation lengths $L_m$ of effective indices $n_m$ and losses per unit lengths $\alpha_m$ (see Fig.\,\ref{Fig03}). The gain sections are modeled by propagation sections exhibiting negative values of $\alpha_m$. Starting from the origin $z=0$, we write the evolution of the mode amplitude $u(z)$ in each section of the cavity. For example, in section $m=0$ ($0\leq z\leq L_0$), the mode amplitude reads just like in eq. (\ref{eq03}). Then the mode amplitude in section $m=1$ ($L_0\leq z\leq L_0+L_1$) must be
\begin{equation}
u(z) = C\sqrt{T_0} \mathrm{e}^{(\mathrm{i}kn_0-\alpha_0/2)L_0}\times\mathrm{e}^{(\mathrm{i}kn_1-\alpha_1/2)(z-L_0)}\ .\label{eq14}
\end{equation}
Once the mode amplitude has been written in all the cavity sections, the normalization factor $C$ can be determined by normalizing the mode amplitude according to eq. (\ref{eq05}), 
 with $L_{\mathrm{tot}}=\sum_m L_m$. To this aim, one also takes into account the self-consistency condition for the laser field:
 \begin{equation}
 1=\prod_{p=0}^M\sqrt{T_p}\exp\left\{ik\sum_{p=0}^Mn_pL_p-\frac{1}{2}\sum_{p=0}^M\alpha_pL_p\right\}\ .\label{eq14N1}
 \end{equation}

The adjoint mode $\varphi(z)$ of $u(z)$ is again calculated by considering the reverse propagation direction (see Supplementary information). 
The adjoint mode is normalized again according to eq. (\ref{eq10})
. Finally, the ENF for mode $u(z)$ is given by eq.\,(\ref{eq12}), leading to:
\begin{eqnarray}
&&K=\ \frac{1}{L_{\mathrm{tot}}^2}\left[L_{0}^{\prime}+\sum_{m=1}^M\left(L_{m}^{\prime}\;\mathrm{e}^{-\sum_{p=0}^{m-1}\alpha_pL_p}\prod_{p=0}^{m-1}T_p\right)\right]\nonumber\\
&&\times\left[\sum_{m=0}^{M-1}\left(L_{m}^{\prime}\;\mathrm{e}^{-\sum_{p=m+1}^{M}\alpha_pL_p}\prod_{p=m}^{M}T_p\right)+L_{M}^{\prime}T_M\right]\;.\label{eq16}
\end{eqnarray}


We can then apply eq.\,(\ref{eq16}) to calculate the phase noises of the two lasers of Fig.\,\ref{Fig01}. The cavity oscillating at 800 nm (see Fig.\;\ref{Fig01}(a)) contains an SOA providing a 26~dB gain, followed by 2 m of fibers, a circulator (3.2 dB of losses per branch, i.e. a total of 6.4 dB)  and Bragg filter (0.6 dB of losses) ensemble, 2 m of fiber, a coupler (3.4 dB losses), 2 m of fiber, a modulator (6 dB losses), again 2 m of fiber followed by an isolator (0.2 dB losses), and finally 2 m of fibers before going back to the SOA entrance. Applying eq. (\ref{eq16}) leads to $K=4$. Similarly, the cavity oscillating at 1550 nm (see Fig.\;\ref{Fig01}(b)) contains an SOA with a 25 dB gain, followed by 2.5 m of fibers , an isolator (0.3 dB losses), 1.5 m of fiber, a coupler (3.5 dB losses), 1.5 m of fiber, a circulator and Bragg filter ensemble (2.7 dB of losses), 2.5m of fiber, a MZM, and finally 103 m of polarization maintaining fiber (1.3 dB losses) before going back to the SOA entrance. Applying eq. (\ref{eq16}) then leads to $K=9$. 

\begin{figure}[htbp]
\centering
\includegraphics[width=0.8\columnwidth]{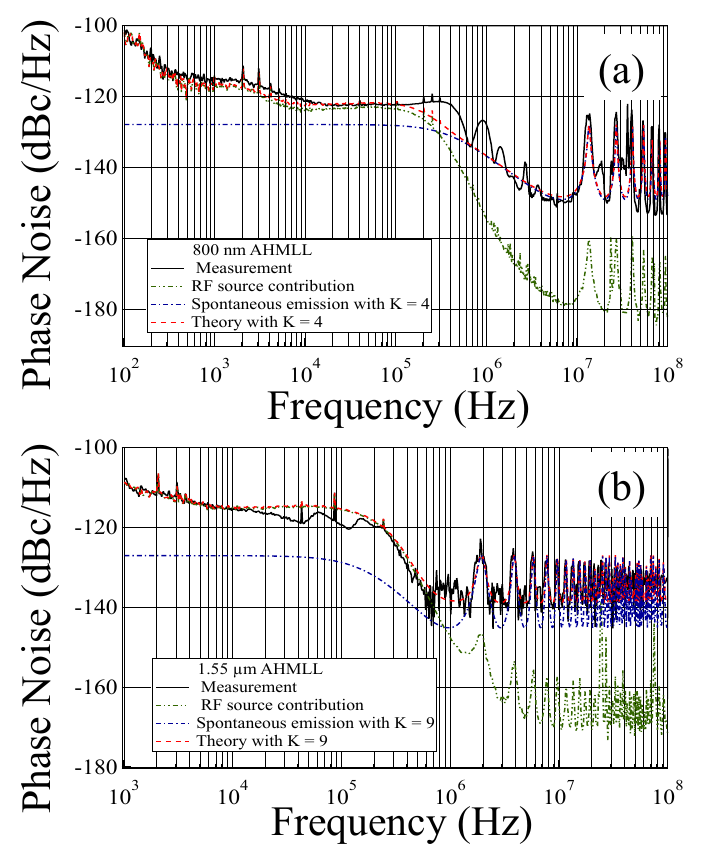}
\caption{(Color online) RF Phase noise spectra of the AHMLL at (a) 800 nm and (b) 1550 nm. Thick full black lines: measurements; Dot-dot-dashed green lines: contribution of the RF source noise; Dash-dotted blue lines: contribution of spontaneous emission; Dashed red lines: theory adding both contributions plus the shot noise level equal to (a) $- 153\;\mathrm{dBc/Hz}$ and (b) $-140\;\mathrm{dBc/Hz}$.}
\label{Fig04}
\end{figure}
When we multiply the Schawlow-Townes linewidth of eq. (\ref{eq02}) by these respective values of $K$ and inject them into eq. (\ref{eq01}), we now obtain the phase noise spectra reproduced in Fig.\,\ref{Fig04}. One can notice that the introduction of the ENF $K$ leads to a strong increase of the calculated noise at high frequencies, especially at the resonance frequencies corresponding to integer multiples of the free spectral range of the cavity. Moreover, the calculated noise spectra are now in excellent agreement with the experimental ones. This illustrates the fact that one cannot reach a good insight in the noise of such systems without taking the non-Hermiticity of the cavity into account.

In conclusion, we have isolated the influence of the non-Hermitian character of the cavity on the phase noise of the train of pulses generated by  actively harmonically mode-locked lasers. This effect has been observed in two different lasers operating at different wavelengths with very different cavity lengths and spatial distributions of  losses and gain. The observed increases in spontaneous emission noise have been shown to be in excellent agreement with the ENF calculated for these non-Hermitian cavities.

This ENF has of course a strong impact on the performances of the clocks and oscillators based on such AHMLLs, and must certainly be taken into account into future designs. But beyond this immediate application, we believe that this work is related to the ultimate performances of devices based on the extreme sensitivity of exceptional points in non-Hermitian optical systems \cite{Longhi2018}. Indeed, large ENFs happen precisely at such exceptional points where eigenvectors become degenerate \cite{vanEijkelenborg1996,Emile1998}. This is exactly the situation that has been predicted to lead to an enhancement of the sensitivity of a ring laser gyroscope \cite{Sunada2017}, without considering that the noise of the system could also be strongly enhanced.

\begin{acknowledgments}
Work partially supported by the  Direction G\'en\'erale de l'Armement (DIFOOL, ANR-15-ASMA-0007-04) and performed in the framework of the joint research lab between TRT and LAC.
\end{acknowledgments}


\begin{thebibliography}{10}
\bibitem{Schawlow1958}
 A.~L.~Schawlow and C. H. Townes, Phys. Rev. \textbf{112}, 1940 (1958).
\bibitem{Jacobs1979}
 S.~F.~Jacobs, Am. J. Phys. \textbf{47}, 597 (1979);  \textbf{49}, 698 (1981).
 \bibitem{Siegman1968}
A.~E. Siegman and R.~Arrathoon, Phys. Rev. Lett. \textbf{20}, 901 (1968).
\bibitem{Manes1971}
K. R. Manes and A. E. Siegman, Phys. Rev. A \textbf{4}, 373 (1971).
\bibitem{SiegmanLivre}
A. E. Siegman, \textit{Lasers,} University Science Books, 1986.
\bibitem{VanExter1995}
M. P. van Exter, S. J. M. Kuppens, and J. P. Woerdman, Phys. Rev. A \textbf{51}, 809 (1995).
\bibitem{Henry1982}
 C. H. Henry, IEEE J. Quantum Electron.  \textbf{QE-18}, 259 (1982).
\bibitem{Henry1986}
 C. H. Henry, J. Lightwave Technol.  \textbf{4}, 298 (1986).
 \bibitem{Siegman1979}
A.~E. Siegman, Opt. Commun. 31, 369 (1979).
  \bibitem{Siegman1989a}
A.~E. Siegman, Phys. Rev. A \textbf{39}, 1253 (1989).
 \bibitem{Siegman1989b}
A.~E. Siegman, Phys. Rev. A \textbf{39}, 1264 (1989).
\bibitem{Mussche1990}
P. L. Mussche and A. E. Siegman, Proc. SPIE \textbf{1376}, 153 (1990).
\bibitem{New1995}
G. H. C. New, J. Mod. Opt. \textbf{42}, 799 (1995).
\bibitem{Grangier1998}
Ph. Grangier and J.-Ph. Poizat, Eur. Phys. J. D \textbf{1}, 97 (1998).
\bibitem{Cheng1996}
Y. J. Cheng, C. G. Fanning, and A. E. Siegman, Phys. Rev. Lett. \textbf{77}, 627 (1996).
\bibitem{vanEijkelenborg1996}
M. A. van Eijkelenborg, Å. M. Lindberg, M. S. Thijssen, and J. P. Woerdman,  Phys. Rev. Lett. \textbf{77}, 4314 (1996).
\bibitem{Brunel1997}
M. Brunel, G. Ropars, A. Le Floch, and F. Bretenaker, Phys. Rev. A \textbf{55}, 4563 (1997).
\bibitem{Petermann1979}
K. Petermann, IEEE J. Quantum Electron. \textbf{QE-15}, 566 (1979).
\bibitem{Hamel1989}
W. A. Hamel and J. P. Woerdman, Phys. Rev. A \textbf{40}, 2785 (1989).
\bibitem{Hamel1990}
W. A. Hamel and J. P. Woerdman,  Phys. Rev. Lett. \textbf{64}, 1506 (1990).
\bibitem{VanderLee1997}
A. M. van der Lee, N. J. van Druten, A. L. Mieremet, M. A. van Eijkelenborg, Å. M. Lindberg, M. P. van Exter, and J. P. Woerdman, Phys. Rev. Lett. \textbf{79}, 4357 (1997).
\bibitem{Emile1998}
O. Emile, M. Brunel, A. Le Floch, and F. Bretenaker, Europhys. Lett. \textbf{43}, 153 (1998).
\bibitem{Hansch2006}
T. W. Hänsch, Rev. Mod. Phys. \textbf{87}, 1297 (2006).
\bibitem{Gerke2016}
S. Gerke, J. Sperling, W. Vogel, Y. Cai, J. Roslund, N. Treps, and C. Fabre, Phys. Rev. Lett. \textbf{117}, 110502 (2016).
\bibitem{Bernhardt2010}
B. Bernhardt, A. Ozawa, P. Jacquet, M. Jacquey, Y. Kobayashi, T. Udem, R. Holzwarth, G. Guelachvili, T. W. Hänsch, and N. Picqué, Nature Photonics \textbf{4}, 55 (2010). 
\bibitem{Fortier2011}
T. M. Fortier, M. S. Kirchner, F. Quinlan, J. Taylor, J. C. Bergquist, T. Rosenband, N. Lemke, A. Ludlow, Y. Jiang, C. W. Oates, and S. A. Diddams, Nature Photonics \textbf{5}, 425 (2011).
\bibitem{Bouchand2016}
R. Bouchand et al., in \textit{Proceedings of the  2016 European Frequency and Time Forum} (IEEE, York, UK, 2016), pp. 352-354, 2016.
\bibitem{Xie2017}
X. Xie et al., Nature Photonics \textbf{11}, 44 (2017).
\bibitem{Mandridis2010}
D. Mandridis, I. Ozdur, F. Quinlan, M. Akbulut, J. J. Plant, P. W. Juodawlkis, and P. J. Delfyett, Appl. Opt. \textbf{49}, 2850 (2010).
\bibitem{Ho1985}
 P.-T. Ho, IEEE J. Quantum Electron. \textbf{QE-21}, 1806 (1985).
 \bibitem{GAF} 
 G. Grynberg, A. Aspect, and C. Fabre, \textit{Introduction to Quantum Optics}, Cambridge, 2010.
 \bibitem{Yao1997}
 X. S. Yao and L. Maleki, Opt. Lett. \textbf{22}, 1867 (1997).
 \bibitem{Matsko2009}
 A. B. Matsko, D. Eliyahu, P. Koonath, D. Seidel, and L. Maleki, J. Opt. Soc. Am. B \textbf{26}, 1023 (2009).
 \bibitem{Matsko2013}
 A. B. Matsko, D. Eliyahu, and L. Maleki, J. Opt. Soc. Am. B \textbf{30}, 3316 (2013).
 \bibitem{Ly2018}
 A. Ly, V. Auroux, R. Khayatzadeh, N. Gutierrez, A. Fernandez, and O. Llopis, IEEE Photon. Tech. Lett. \textbf{30}, 1313 (2018).
 \bibitem{Hjelme1992}
 D. R. Hjelme and A. R. Mickelson, IEEE J. Quantum Electron. \textbf{QE-28}, 1594 (1992).
 \bibitem{Quinlan2008} 
 F. J. Quinlan,  PhD thesis, University of Central Florida, Orlando, Florida, 2008.
 \bibitem{Quinlan2009} 
 F. Quinlan, S. Ozharar, S. Gee, and P. J. Delfyett, J. Opt. A: Pure Appl. Opt. \textbf{11}, 103001 (2009).
 \bibitem{Alouini2002}
 M. Alouini, F. Bretenaker,  M. Brunel, D. Chauvat, O. Emile, N. D. Lai, A. Le Floch, G. Ropars, and M. Vallet, Acta Physica Polonica \textbf{101}, 7 (2002).
 \bibitem{Longhi2018}
 S. Longhi, Opt. Lett.  \textbf{43}, 2929 (2018).
 \bibitem{Sunada2017}
 S. Sunada, Phys. Rev. A \textbf{96}, 033842 (2017).
\end{thebibliography}
\end{document}